\documentclass[letter,tradiabstract]{aa} 
\usepackage{graphicx}
\usepackage{txfonts}
\usepackage[authoryear]{natbib}
\bibpunct{(}{)}{;}{a}{}{,}

\newcommand{\HI}{{\ion{H}{i}}}

\newcommand{\kms}{$\,$km$\,$s$^{-1}$}

\newcommand{\mJybeam}{mJy beam$^{-1}$}

\def\emph#1{{\sl #1}} 
\newcommand{\ltsima} {$\; \buildrel < \over \sim \;$} 
\newcommand{\gtsima} {$\; \buildrel > \over \sim \;$}
\newcommand{\lta} {\lower.5ex\hbox{\ltsima}} 
\newcommand{\gta}{\lower.5ex\hbox{\gtsima}}

\begin{document} %
\title{A circumnuclear disk of atomic hydrogen in  Centaurus A} 
\author{R. Morganti
\inst{1,2}, T. Oosterloo\inst{1,2}, C. Struve\inst{1,2} \and L.
Saripalli\inst{3,4} }

\institute{Netherlands Institute for Radio Astronomy, Postbus 2, 7990 AA,
Dwingeloo, The Netherlands \and Kapteyn Astronomical Institute, University of
Groningen, Landleven 12, 9747 AD, Groningen, The Netherlands \and Raman Research
Institute, CV Raman Avenue, Sadashivanagar, Bangalore 560080, India \and CSIRO,
Australia Telescope National Facility, PO Box 76, Epping NSW 1710, Australia }
\date{\today}

\abstract {We present new observations, performed with the Australia Telescope
Compact Array,  of the \HI\ absorption in the central regions of Centaurus~A.
For the first time, absorption is detected against the radio core at velocities
blueshifted with respect to the systemic velocity. Moreover, the  data  show
that the nuclear redshifted absorption component is broader than reported
before. With these new results, the kinematics of the  \HI\ in the inner regions
of Cen A appears very similar to that observed in emission for the molecular
circumnuclear disk. This suggests that the central \HI\ absorption is not, as was previously
claimed, evidence of  gas infall into the AGN, but instead is due to a cold,
circumnuclear disk.}
 
\maketitle 
%

\section{Introduction}
In the central regions of galaxies with an Active Galactic Nucleus (AGN), gas
can play different, sometimes competing roles. This gas is considered to be
essential for fuelling the AGN and for turning a dormant black hole into an
active one. On the other hand, the gas can also be expelled from these regions
as a result of the release of energy from the active nucleus, so that fuelling
and star formation are quenched. Moreover, some of the nuclear gas can be
arranged in regular  structures.  Such circumnuclear disks, under certain
geometries, can hide the active nucleus from our direct view and can play an
important role in explaining the apparent differences between the various types
of AGN. The conditions in these central regions are clearly harsh, but
nevertheless they are such that not only very dense or highly ionised clouds,
but also atomic and molecular gas can survive.

It has been suggested that neutral hydrogen falling onto the central black hole
is a possible mechanism for fuelling  an AGN. For many years, the available data
seemed to indicate that if \HI\ absorption is detected against the core of a
radio-load AGN, it is seen at velocities redshifted with respect to the systemic
\citep[see e.g.][]{vg89}. Although other factors, such as non-circular motions, can explain redshifted absorption in a single galaxy \citep[see e.g.][]{gb07}, the strong predominance of redshifted absorption has been
interpreted as  evidence for gas falling into the nucleus. However, it turns out that the relatively narrow observing bands used in
the past have led to an incomplete picture \citep[see e.g.][]{mor05}.  Many
cases of blueshifted \HI\ absorption against the radio-core are now known
\citep{ver03,mor05} and as a result, the evidence for nuclear \HI\ absorption to
be more often redshifted than blueshifted, and hence for gas infall fuelling the
AGN, has disappeared in recent years. A particularly well-known example of
possible \HI\ infall is the detection of redshifted absorption against the
nucleus of the closest radio galaxy, Centaurus A\footnote{At the assumed
distance of Cen A of 3.5 Mpc, 1 arcsec corresponds to $\sim 17$ pc, or
equivalently, 1 arcmin to 1 kpc} \citep{vdh83}.  This case is important because
if it would indeed be due to gas falling into the AGN, the proximity of Cen A
would allow a detailed study of the infall and related processes.

In this Letter we present new observations  of the \HI\ absorption in Cen A that
suggest a different interpretation of what causes the \HI\ absorption against
the core of this galaxy.   The broad observing band used has allowed us to
detect, for the first time,  {\sl blueshifted} \HI\ absorption against the core
of Cen A. This obviously complicates the interpretation of the \HI\ absorption
as infall. Below, we discuss the various possibilities and  we conclude that it
is evidence that the \HI\ absorption is caused by a cold, circumnuclear disk and
not by infall into the AGN.

\section{The ATCA Data}

Here, we only give a brief summary on the new observations of Cen A, a full
description, including results on the large-scale gas disk, will be presented in
a forthcoming paper (Struve et al., in prep). The observations were done using
the Australia Telescope Compact Array (ATCA) in three standard 750-m, 1.5-km and
6-km configurations, and were carried out on  12, 14 and 20 April 2005 (12 h
in each configuration). The choice of the  configurations used was made to
obtain good spatial resolution, but  at the same time to avoid to include too
many too short baselines. Given the very strong, very extended continuum
emission of Cen A, bandpass calibration is very difficult for short baselines (see below). A 16-MHz band  with 512 channels was used, equivalent to a velocity
range of $\sim$3000 \kms. This is much broader than used in earlier observations
\citep{vdh83,sar02}. The centre of the band was set  to $V_{\rm hel}=304$ \kms\
and the velocity resolution is $13.2$ \kms, after Hanning smoothing.
The broad band was chosen to make sure that the full width of the central \HI\
absorption would be sampled if broad signals would appear (as have been seen in
other radio galaxies, see \cite{mor05}). PKS~1934--638 was used as bandpass
calibrator. Short observations of about 10 minutes were done every hour on the
phase calibrator PKS 1315--46.  The data reduction was carried out using the
MIRIAD package \citep{sault95}.

The extremely strong radio continuum of Cen~A ($> 100$ Jy on the shortest
baseline) makes the bandpass calibration the most delicate part of the
calibration. We have used the same technique  applied by \cite{oost05} on
earlier observations of Cen A. This involves smoothing the bandpass calibration
obtained from PKS~1934-638 with a box-car filter 15 channels wide.   This
smoothing effectively increases the flux level of PKS 1934--638 to about 60 Jy
which is higher than the detected flux on almost all baselines of Cen A. This
procedure is allowed because the features in the instrumental bandpass are
fairly broad in frequency.  This smoothed bandpass correction is applied to the
{\sl unsmoothed} data of Cen~A.The resulting spectral dynamic range is better
than 1:10000 on the shortest baselines.

The subtraction of the continuum was done using the task UVLIN, making a second order fit to the line-free channels of each visibility record and subtract this fit from the spectrum. In order to improve on the phase calibration obtained from
PKS 1315--46, frequency independent self-calibration was  performed taking
advantage of the strong nuclear \HI\ absorption  at some velocities. The final
cube was made using the combined datasets. In this Letter we will use the
high-resolution cube obtained with uniform weighting and including the 6-km
antenna. The lower resolution cubes will be discussed in Struve et al.\ (in
prep).  The restoring beam is $8.1\times 6.8$ arcsec with ${\rm PA} =
-12.3^\circ$. The rms noise per channel is $\sim 1.3$ \mJybeam. The continuum
image   obtained from the line-free channels   is shown in Fig.\ 1.

\begin{figure} 
\centering 
\includegraphics[width=6.2cm]{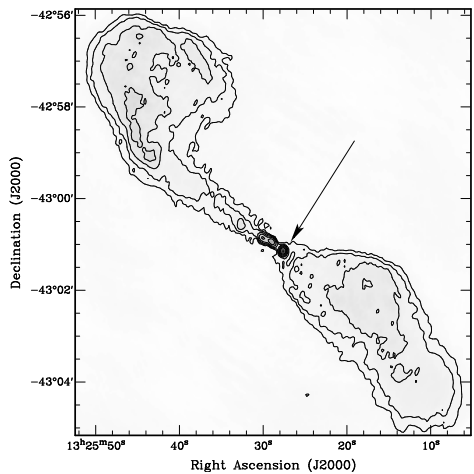} 
\caption{The
continuum image of Cen A obtained from the line-free channels. The absorption discussed here is against the central bright source (indicated by the arrow) } 
\end{figure} 
%

\section{Broad nuclear  \HI\ absorption}

As expected, we detect \HI\ in emission and, against the strong continuum, in
absorption.   A full description of these data will be given in Struve et al.\
(in prep). In this letter we focus  on the \HI\ absorption detected against the
{\sl very central regions}.

Both deep \HI\ absorption near the systemic velocity and  fainter redshifted
absorption against the radio core had been detected before by \citet{vdh83} and
\citet{sar02}. Our data show, of course, the deep absorption component near the
systemic velocity. However, the new and interesting result is that in our new
data, the fainter absorption against the nucleus  {\sl covers a much larger
velocity range compared to previous observations}. This larger width of the
fainter component is   due to both the fact that the redshifted absorption
extends to more extreme velocities and to a newly detected blueshifted
component. The absorption spectrum is shown in Fig.\ 2.   The total velocity
width is about 400 \kms. The bandwidth used by \citet{vdh83} was 300 \kms\ while
that used by \citet{sar02}  only 150 \kms. It is therefore clear that it is due
to the limited bandwidth used  that only part of the redshifted \HI\ absorption 
 was detected in those observations and that  the blueshifted absorption was not
detected at all.

The nuclear \HI\ absorption appears asymmetric with respect to the systemic
velocity (542 \kms\ \citet{vg90}, covering (heliocentric) velocities from $\sim$$400$ \kms\ (about
$-140$ \kms\ blueshifted compared to  systemic) up to $\sim$$800$ \kms\ (i.e.\
about +260 \kms\ redshifted relative to  systemic).  The peak absorption has an
optical depth of $\tau = 0.83$. The column density of the deeper absorption
component is $N_{\rm \HI} \sim 2.7\times 10^{19}$ $T_{\rm spin}$ cm$^{-2}$. This
column density is in good agreement with what derived by \citet{vdh83}.  One
should note that the spin temperature $T_{\rm spin}$ for \HI\ absorption
occurring close to an AGN is a very uncertain parameter.  Another uncertainty in
$N_{\rm HI}$ is that above we have  assumed that the absorption covers the
entire underlying continuum source (see also below).

\subsection{A circumnuclear \HI\ disk?}

One of the central questions is of course what structure in Cen A is causing
the faint, broad nuclear \HI\ absorption that we see. The detection of
blueshifted absorption obviously complicates the interpretation that it is due
to infall.  We can exclude that the broad, fainter absorption is produced by
gas located at large (kpc) distance from the nucleus.  Figure 2 shows that the gas in the
large-scale disk is on (almost) circular orbits and that absorption  due to the large-scale disk against the radio core  is near the systemic velocity. Projected on the centre, the large disk shows some spread in velocity (due to  non-circular motions) but this spread is much smaller than required to explain the broad central absorption.  
If some of the absorption would instead occur against the
inner continuum jet, the fact that the jet is more or less aligned with the
minor axis of the large-scale gas disk means that we would still expect this
absorption to be mainly at the systemic velocity. The deep absorption
component near the systemic velocity is most likely due to the large-scale gas
disk seen in front of the core and jet, but the fainter, broad absorption
cannot be caused by it.

One possibility for explaining the nuclear absorption at velocities away from
systemic is to assume that the large-scale gas disk is quite thick. In the
models of \cite{Eck90}, the redshifted nuclear absorption components are 
explained by high-latitude clouds of the large-scale disk at radii of $\sim$500
pc. However, such models  fail to explain the  simultaneous occurrence of
bluesifted and redshifted absorption. This means that  in order to reproduce
the observed width of the broad, shallow absorption, large radial motions, both
inward and outward, would  be required to exist in the large-scale disk and such
motions are not observed.

\begin{figure*} \centering \includegraphics[width=16.9cm]{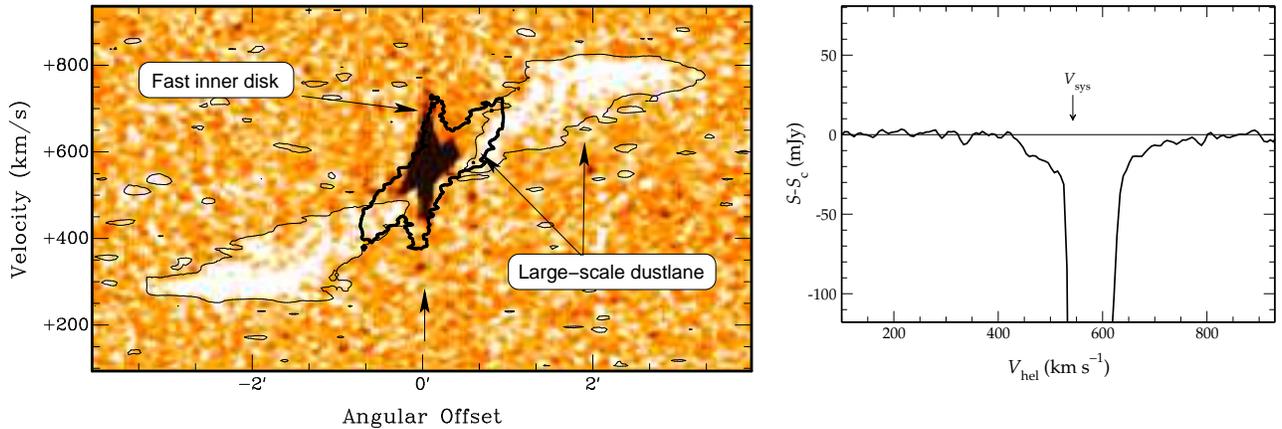}
\caption{{\sl left:} Position-velocity plot of the \HI\ (grey-scale and thin contours) and
superimposed the CO emission (thick contours; from \citet{liszt01}, taken
along position angle 139$^\circ$.  The gray-scale represents the
high-resolution \HI\ data (beam 8 arcsec) while the thin contour is from the
same dataset smoothed to 15 arcsec.  Note
 that the CO observations do not extend beyond a radius of about 1 arcmin.
{\sl right:} The absorption profile against the central radio core (indicated by the arrow in tthe left panel)  showing the blue- and redshifted wings of the absorption profile } \end{figure*} %

An entirely different possibility is that  the broad, shallow absorption is
caused by a circumnuclear \HI\  disk. If so, one of the prime cases for cold gas
falling into an AGN would disappear.

The presence of  a cold circumnuclear disk with a size of roughly 100 parsec has
been suggested by a number of studies \citep[see e.g.][and references
therein]{isr90}. The central regions of Cen~A are highly obscured in the
optical, but they have been studied in detail using observations at longer
wavelengths. Ionised and molecular gas was observed in the K-band using both
long-slit \citep{mar01}  and integral-field spectroscopy \citep{kra07,neu08}.
The most recent  observations \citep{neu08}  show that in the central few
arcseconds (i.e.\ tens of pc) very different morphologies are found between the
high- and low-ionisation lines. The  highly ionised gas shows a strong
kinematical influence of the jet. On the other hand,  the molecular gas (H$_2$)
appears to be in a regularly rotating disk, influenced only by gravity. The
velocity channel maps \citep[Fig.\ 8 in ][]{neu08} show the kinematics of this
inner molecular disk, with velocities ranging from $\sim$$-230$ \kms\ to
$\sim$$230$ \kms\ compared to systemic.  Part of this velocity range is due to
the velocity dispersion of the gas, whose origin is not known, the rest is due
to rotation. The H$_2$ kinematics are best modelled by a tilted-ring model of a
warped gas disk that, in the inner part, appears quite face on (34$^\circ$) and
with position angle  of the kinematic major axis of $\sim 155^\circ$.

Other evidence for a circumnuclear molecular disk comes from CO observations
\citep{Eck90,isr90,Ryd93}.  In particular, JCMT CO $J$=3-2 observations
\citep{liszt01} have been explained by a disk of radius 168 pc seen nearly
edge-on (in position angle 138.5$^\circ$ on the sky,
therefore perpendicular to the base of the radio jet). In these observations -
that have a resolution of 14 arcsec - this disk is barely resolved. For both the
CO and the H$_2$ data the kinematic modelling is somewhat uncertain.
Nevertheless, the difference in the inclination derived for the inner H$_2$
disks and the somewhat larger CO structure suggests that it is  heavily warped
at all spatial scales.

In an attempt to see whether  the broad \HI\ absorption could be related to this
circumnuclear disk, we have compared the CO $J$=3-2 data (kindly provided by H.\
Liszt) and the \HI\ data cubes. Figure 2 shows the position-velocity plot
obtained along a line (position angle 139$^\circ$) through the nucleus  of Cen
A.  The gray-scale and the thin contours represent the \HI\ data while the CO
data is given by the thick contours. Note that the field of view of the CO
observations was only $2\times2$ arcmin so the CO information does not extend
beyond  1 arcmin radius.

The CO data clearly show two kinematical components: a large component with a
relatively shallow velocity gradient and a smaller, inner component with a steep
velocity gradient. The large CO component corresponds to the large-scale gas
disk associated with the dustlane, while the inner component is the
circumnuclear disk modelled by \cite{liszt01}. The comparison with the \HI\
clearly suggests that both the outer (in emission, and part in
absorption) and the inner component (in absorption) are seen also in \HI. Figure
2 illustrates the point made earlier that the kinematics of the large disk is
clearly such that it cannot explain the large velocity range seen in the very
centre.

For the inner part, Fig.\ 2 shows that the broad, inner \HI\ absorption
corresponds very well with the fast rotating, inner CO disk.  In
particular, the redshifted \HI\ absorption covers the same velocity range as the
CO, suggesting that it is not associated with infalling gas. It
is also clear that the \HI\ and CO show differences: the extent of the CO, both
spatially and in the blueshifted velocities, is larger than of the inner \HI.
Such differences can be expected given that the characteristics of the inner
\HI\ absorption is set by the size of the background continuum emission, {\sl
not} by the \HI\ structure itself.  With the spatial resolution of our \HI\ data
(8 arcsec), the inner \HI\ absorption is unresolved while the total extent of
the CO disk is about 20 arcsec. If the background continuum against which we see
the \HI\ absorption is smaller than the circumnuclear CO disk (see below), this
would explain the apparent difference in size and velocity range.

\subsection{The nature of the absorbed continuum}

As already eluded to above, for the interpretation of the \HI\ absorption, it is
important to understand what the continuum background could be. It turns out
that it is not quite obvious to unravel this. It is natural to assume that the
strongest absorption is against the very inner radio core. However, this is
unlikely to be the case. The radio continuum emission on the sub-kpc scale has
been studied using high-frequency VLA (about 0.5 arcsec resolution at 8.4~GHz,
Hardcastle et al.\ 2003) as well as VLBI observations at frequencies 2.3~GHz and
higher, with very high spatial resolution \citep[a few milliarcsec resolution at
2.3~GHz;][]{jones96,ting98}.  The highest frequency data show that, on parsec
and sub-parsec scales, the radio continuum in the central region is dominated by
a core-jet structure. However, the  VLBI data also show that the core has
a strongly inverted spectrum and that it is actually {\sl not visible at 2.3 GHz}.
This suggests that the very innermost part of the radio source (up to $\sim
0.4-0.8$ pc) is seen through a disk or torus of ionised gas which is opaque at
lower frequencies due to free-free absorption \citep{jones96}. The above means
that  the radio core is unlikely to be detected also at 1.4~GHz and it is not the core itself against which the 21-cm absorption can be
occurring.

Given that, at 1.4 GHz, the radio core is not seen, the radio continuum on the
milli-arcsec scale appears concentrated in the jet and counter-jet. Given the
circumnuclear disk structure discussed above, with a PA close to perpendicular
to the jet axis, the fact that the main jet is coming toward us means that it is
likely to be in front of the circumnuclear disk and that it cannot cause the
absorption. One possibility would be then that the \HI\ absorption is against
the {\sl counter-jet}.  However, given the small opening angle of the jet this
can be excluded as most of the absorption would be seen at velocities close to
systemic.

An interesting clue is that  extra flux (at least 50 mJy) is clearly present on the short
baselines in the 1.4-GHz VLBI experiments  (Tingay priv.\
comm.), indicating  the presence of
large-scale  structures that are not evident in the images. This structure would be still well inside the 1 arcsec ($\sim$18 pc)
imaged at the highest VLA resolution (for 21 cm). It is difficult to quantify
the  amount of flux on larger scales from a comparison of VLBI data with
lower-resolution data due to the variability of the nuclear regions
\citep{rom97,Abr07}  because simultaneous observations are needed. It is
therefore unclear whether this extra component  can cause the absorption.

If, however,  this would be the case, we have a situation similar to that
suggested for Cygnus~A (Conway \& Blanco 1995, Conway 1999). In this source, the
\HI\ opacity peaks off the jet, on the counter-jet side, with maximum opacities
occurring north and south of the jet axis.  The counter-jet of Cyg A alone is
too narrow to explain the broad width of the absorption. Therefore, Conway
(1999) suggested that a more diffuse component could be at the origin of the
absorption. This component could be due to thermally emitting ionised gas
evaporated from the inner edge of a torus or disk.  Also in Mrk~231 evidence of
\HI\ absorption against a diffuse component has been found (Carilli et al.\
1998). In this object, they identify the \HI\ and the radio continuum disk as
the inner part of the molecular disk seen on the larger scale.

If the scenario described above is correct, we can make some remarks about the
conditions of the \HI\ material and whether the location of the \HI\ and of
the molecular gas are what expected from the physical models of the
circumnuclear disks/tori. In this case, the column density of the \HI\ is
going to be much larger than  estimated above, because the absorption will
likely cover only a small fraction of the sub-arcsec continuum. For example,
if the absorption is only against the counterjet-side, and assuming the
jet/counterjet ratio from Jones et al.\ (1996) of 4 - 8, the column density is
at least a factor four higher. This factor would be even higher in the case of
the absorption against the more diffuse component. In addition, in a
circumnuclear structure, $T_{\rm spin}$ is likely to be a few $\times 10^3$ K,
as the \HI\ is affected by the central AGN (Maloney et al.\ 1996). Therefore the
column density could easily be as high as $5 \times 10^{23}$
cm$^{-2}$. Interestingly, this is a similar column density as derived from the
hard X-ray spectrum of Cen A that is fitted with an heavily absorbed power-law
model with a column density of $\sim$$10^{23}$ cm$^{-2}$ (Evans et al.\ 2004).

Physical models of the circumnuclear disks/tori have been investigated by
Maloney et al.\ (1994, 1996 and refs therein).  They identify an effective
ionisation parameter that determines whether, given the distance from the
nucleus, the material is atomic or molecular.  This effective ionisation
parameter is $\xi_{\rm eff} = L_{\rm X} N^{-0.9}_{22}n^{-1}r^{-2}$, where
$L_{\rm X}$ is the luminosity of $>2$ keV X-rays, $n$ is the  gas
density, $r$ is the distance to the nucleus, $N_{22}$ is the total column
density in units of $10^{22}$ cm$^{-2}$.  For $\xi_{\rm eff} > 10^{-3}$, the
gas will be largely atomic, otherwise mainly molecular.  For Cen~A, the X-ray
luminosity is $L_{\rm X} = 5\times 10^{41}$ erg/s (Evans et al.\ 2004) and the
X-ray column density is $10^{23}$ cm$^{-2}$. A rough estimate of the density
can be derived from the \HI\ column density if we assume that the region of
absorption is comparable to the region with free-free absorption (that covers
the innermost $\sim 0.4-0.8$ pc).  This gives a density of $ 1-2 \times 10^5$
cm$^{-3}$.  These values give a radius of 1-3 pc for
$\xi_{\rm eff} = 10^{-3}$.  Thus, we may hypothesise that the central region
of Cen~A hosts a gaseous disk composed of multiple phases where the inner part
(but outside the fully ionised region) is mainly atomic and after that
molecular.

\section{Conclusions}

Using new,  broad-band, \HI\ observations of Cen~A we have shown that the
nuclear absorption is broader than previously known and that it has a
blueshifted component.  Using various arguments, but in particular the
comparison with the molecular gas data, we have shown that the \HI\ absorption
is likely to be caused by a cold, circumnuclear disk and that it does not
constitute direct evidence of  gas infall into the AGN.

Sensitive VLBI observations are now needed to further explore the
characteristics of the nuclear \HI\ as well as the picture proposed in this
paper and, in general, to learn more about the central regions of this
fascinating, nearby AGN.

\begin{acknowledgements} We are most grateful to Harvey Liszt for providing us
the CO cube that has been crucial in this work.  Based on observations with the
Australia Telescope Compact Array (ATCA), which is operated by the CSIRO
Australia Telescope National Facility. This research was supported by the EU
Framework 6 Marie Curie Early Stage Training programme under contract number
MEST-CT-2005-19669 ESTRELA.
\end{acknowledgements}

\end{document}